\documentclass[preprintnumbers, showpacs, floatfix,onecolumn,preprintnumbers, letterpaper, superscriptaddress,nofootinbib]{revtex4}
\usepackage{amsfonts}
\usepackage{amsmath}
\usepackage{amssymb,epsf}
\usepackage{latexsym}
\usepackage{graphicx,epsfig}
\usepackage{amssymb}
\usepackage{float}
\usepackage{subfigure}
\usepackage{epstopdf}
\usepackage[colorlinks=true,citecolor=blue,linkcolor=blue,urlcolor=black]{hyperref}
\usepackage{dcolumn}
\usepackage{psfrag}
\usepackage{wrapfig}
\usepackage{makeidx}
\usepackage{epsf}

\begin{document}

\title{One-dimensional backreacting holographic $p$-wave superconductors}
\author{Mahya Mohammadi}
\affiliation{Physics Department and Biruni Observatory, Shiraz University, Shiraz 71454,
Iran}
\author{Ahmad Sheykhi}
\email{asheykhi@shirazu.ac.ir}
\affiliation{Physics Department and Biruni Observatory, Shiraz University, Shiraz 71454,
Iran}
\affiliation{Research Institute for Astronomy and Astrophysics of Maragha (RIAAM), P. O.
Box: 55134-441, Maragha, Iran}
\author{Mahdi Kord Zangeneh}
\email{mkzangeneh@scu.ac.ir}
\affiliation{Physics Department, Faculty of Science, Shahid Chamran University of Ahvaz,
Ahvaz 61357-43135, Iran}

\begin{abstract}
We analytically as well as numerically study the properties of
one-dimensional  holographic $p$-wave superconductors in the
presence of backreaction. We employ the Sturm-Liouville eigenvalue
problem for the analytical calculation and the shooting method for
the numerical investigations. We apply the AdS$_{3}$/CFT$_{2}$
correspondence and determine the relation between the critical
temperature $T_{c}$ and the chemical potential $\mu $ for
different values of mass $m$ of charged spin-$1$ field $\rho _{\mu
}$ and backreacting parameters. We observe that the data of both
analytical and numerical studies are in good agreement. We find
out that increasing the backreaction as well as the mass parameter
cause the greater values for ${T_{c}}/{\mu }$. Therefore, it makes
the condensation harder to form. In addition, the
analytical and numerical approaches show that the value of the critical exponent $%
\beta $ is ${1}/{2}$ which is the same as in the mean field
theory. Moreover, both methods confirm the exhibition of a second
order phase transition.
\end{abstract}

\maketitle

\section{Introduction}
In $1911$ Heike Kamerlingh Onnes discovered that the electrical
resistance of mercury completely disappeared at temperatures a few
degrees above absolute zero \cite{Dahi}. The phenomenon became
known as superconductivity. He was also awarded the the Nobel
Prize in Physics $1913$ for his investigations on the properties
of matter at low temperatures which led, inter alia, to the
production of liquid helium. Since the discovery of Onnes, the
studies on the superconductors have been became an active field of
research and a lot of papers have been appeared in the literatures
to explain the mechanism of superconductivity. The aim was to
explain the zero resistance of the materials from microscopic
point of view. The great step in this direction put forwarded in
$1957$ by John Bardeen, Leon Neil Cooper and John Robert
Schrieffer who described superconductivity as a microscopic effect
caused by a condensation of Cooper pairs into a boson-like state.
They also awarded the Nobel Prize in physics $1972$ for their
jointly developed theory of superconductivity, usually called the
BCS-theory. The BCS theory, however, requires only that the
potential be attractive, regardless of its origin. In the BCS
framework, superconductivity is a macroscopic effect which results
from the condensation of Cooper pairs. It was the first
widely-accepted theory that explained superconductivity at low
temperatures. Based on this theory superconductivity occurs
because of condensation of Cooper pairs (including electrons with
different spins and momenta) at low temperature. According to
angular momentum of Cooper pairs, we can classify
superconductors as $s$-wave $(\ell =0)$, $p$-wave $(\ell =1)$, $d$-wave $%
(\ell =2)$ etc \cite{alkac}. Since the Cooper pairs are decoupled
at higher temperatures, the BCS theory has argued to be inadequate
to fully explain the mechanism of high temperature
superconductivity \cite{BCS57}. In order to shed some light on the
problem of high temperature superconductivity the Anti de
Sitter/Conformal Field Theory (AdS/CFT) correspondence was argued
to taken into account \cite{Maldacena,H08}. AdS/CFT duality is a
duality that relates the strong coupling conformal field theory
living on the boundary in $d$-dimensions to a weak coupling
gravity in ($d+1$ )-dimensional spacetime in the bulk. Through
AdS/CFT, each quantity in the bulk has a dual on the boundary
\cite{Maldacena,H08,G98,W98,HR08,R10}. In $2008$, Hartnoll et al.
proposed a holographic $s$-wave superconductor model based on the
gauge/gravity duality \cite{H08}. In his holographic model,
Hartnoll assumed that there is a phase transition from a black
hole with no hair (normal phase) to a hairy one (superconducting
phase) below the critical temperature. Through this process, the
system faces with spontaneous $U(1)$ symmetry breaking. The
studies on the holographic superconductors have arisen a lot of
attentions in the past decade (see e.g.
\cite{H09,Hg09,H11,Gu09,HHH08,JCH10,SSh16,SH16,cai15,SHsh(17),
Ge10,Ge12,Kuang13,Pan11,Wang6,CAI11, SHSH(16),shSh(16),Doa,
Afsoon, cai10,yao13,n3,n4,n5,n6,Gan1} and references therein).

The holographic $p$-wave superconductors can be studied by
condensation of a charged vector field in the bulk which is the
dual of a vector order parameter in the boundary which can also be
considered as the condensation of a $2$-form field in the
boundary. For this type of holographic superconductor, the
formation of vector hair below the critical temperature is
observed. Various models of holographic $p$-wave superconductors
have been proposed. In \cite{Gubser} a $p$-wave superconductors
proposed by using an SU(2) Yang-Mills field in the bulk and one of
the gauge degrees of freedom which is dual to spin-$1$ order
parameter in the field theory. Also, the $p$-wave type of
superconductivity may arisen by the condensation of a $2$-form
field \cite{Donos} and a massive spin-$1$ vector in the bulk \cite
{Caip,cai13p}. The holographic $p$-wave superconductors have been
widely investigated in the literatures
(e.g.\cite{Roberts8,zeng11,cai11p,pando12,momeni12p,gangopadhyay12,chaturverdip15}
).

on the other side, holographic superconductors have also been
explored when the bulk spacetime is a three dimensional black
hole. The Einstein field equations admit a three dimensional
solution known as BTZ (Bandos-Teitelboim-Zanelli) black holes. BTZ
black holes
have a crucial effect on the several improvement in string theory \cite%
{Car1,Ash,Sar,Wit1,Car2}. The corresponding superconductor living
on the boundary of BTZ black hole is one dimensional. Using the
probe brane construction, the holographic $p$-wave superconductors
were investigated in \cite{Bu}. One dimensional holographic
$p$-wave superconductors coupled to a massive complex vector field
and in the probe limit were explored in \cite{alkac}. It was
argued that below a certain critical temperature,  there is a
formation of a vector hair around the black hole \cite{alkac}. It
is worth noting that in order to analyze one-dimensional
holographic superconductor on the boundary of the three
dimensional spacetime, one needs to apply the AdS$_{3}$/CFT$_{2}$
\cite{Wit2}. One-dimensional holographic $s$-wave and $p$-wave
superconductors were investigated analytically as well as
numerically from different point of view (see e.g.
\cite{Wang,chaturvedi,L12,momeni,peng17,lashkari,hua,yanyan,yan,kord,bina,mahya}).
All investigations on the ($1+1$)-dimensional holographic $p$-wave
superconductors are restricted to the case where the vector and
gauge fields do not back react on the background geometry. In the
present work, we would like to extend the study on the holographic
$p$-wave superconductors by considering the effects of the vector
and gauge fields on the background of spacetime and disclose the
effects of the backreaction on the properties of superconductor.

We shall employ the Sturm-Liouville eigenvalue problem for the
analytical calculation and the shooting method for the numerical
investigations. For each method, the relation between critical
temperature and chemical potential as well as critical exponent
are investigated. We shall also compare the analytical results
with the numerical data.

This paper is outlined as fallow. In section \ref{basic}, we
present the basic field equations and the boundary conditions of
the ($1+1$)-dimensional backreacting holographic $p$-wave
superconductors. In section \ref{phase}, by using the
Sturm-Liouville variational method, we obtain a relation between
the critical temperature and the chemical potential. We also apply
the shooting method and study the problem numerically and confirm
that the analytical results are compatible with the numerical
data. In section \ref{Crit}, we calculate the critical exponent
both analytically and numerically. The last section is devoted to
closing remarks.
\section{Basic Field Equations and boundary conditions\label{basic}}
As we mentioned our study is based on the AdS$_{3}$/CFT$_{2}$
duality. Due to this model, we have a spontaneous local/global
$U(1)$ symmetry breaking in the bulk/at the boundary. The action
which can describe a charged massive
spin-$1$ field $\rho _{\mu }$ with charge $q$ and mass $m$ into ($2+1$%
)-dimensional Einstein-Maxwell theory with a negative cosmological
constant is given by
\begin{eqnarray}
S &=&\frac{1}{2\kappa ^{2}}\int d^{3}x\sqrt{-g}\left( R+\frac{2}{l^{2}}%
\right) +\int d^{3}x\sqrt{-g}\mathcal{L}_{m},  \notag \\
\mathcal{L}_{m}&=&-\dfrac{1}{4}F_{\mu \nu }F^{\mu \nu }-\frac{1}{2}%
\rho _{\mu \nu }^{\dagger }\rho ^{\mu \nu }-m^{2}\rho _{\mu
}^{\dagger }\rho ^{\mu }+iq\gamma \rho _{\mu }\rho _{\nu
}^{\dagger }F^{\mu \nu }, \label{act}
\end{eqnarray}%
where $g$, $R$ and $l$ are the metric determinant, Ricci scalar
and AdS radius, respectively. $\kappa ^{2}=8\pi G_{3}$ in which
$G_{3}$ characterizes the $3$-dimensional Newton gravitation
constant in the bulk. Also, by considering $A_{\mu }$ as the
vector potential, the strength of Maxwell field reads $F_{\mu \nu
}=\nabla _{\mu }A_{\nu }-\nabla _{\nu }A_{\mu }$. In addition,
$\rho _{\mu \nu }=D_{\mu }\rho _{\nu }-D_{\nu }\rho _{\mu }$ where
$D_{\mu }=\nabla _{\mu }-iqA_{\mu }$. A nonlinear interaction
between $\rho _{\mu }$ with $\gamma $ (the magnetic moment) and
$A_{\mu }$ is described by the last term in the above action.
Since we consider the case without external magnetic field, this
term plays no role.

We obtain the equations of motion for matter and gravitational
fields by varying
the action (\ref{act}) with respect to the metric $g_{\mu \nu }$, the gauge field $A_{\mu }$ and the vector field $%
\rho _{\mu }$. We find
\begin{eqnarray}
\frac{1}{2\kappa ^{2}}\left[ R_{\mu \nu }-g_{\mu \nu }\left( \frac{R}{2}+%
\frac{1}{l^{2}}\right) \right] &=&\frac{1}{2}F_{\mu \lambda
}F_{\nu }{}^{\lambda }+\frac{1}{2}\mathcal{L}_{m}g_{\mu \nu
}+\frac{1}{2}\left[ \rho ^{\dagger }{}_{\mu \lambda }\rho _{\nu
}^{\lambda }+m^{2}\rho ^{\dagger }{}_{\mu }\rho _{\nu }-i\gamma
qF_{\nu }^{\lambda }\left( \rho _{\mu }\rho ^{\dagger }{}_{\lambda
}-\rho ^{\dagger }{}_{\mu }\rho _{\lambda }\right)
+\mu \leftrightarrow \nu \right] .  \notag \\
&&  \label{Eein}
\end{eqnarray}%
\begin{equation}
\nabla ^{\nu }F_{\nu \mu }=iq\left( \rho ^{\nu }\rho ^{\dagger }{}_{\nu \mu
}-\rho ^{\nu \dagger }\rho _{\nu \mu }\right) +iq\gamma \nabla ^{\nu }\left(
\rho _{\nu }\rho ^{\dagger }{}_{\mu }-\rho ^{\dagger }{}_{\nu }\rho _{\mu
}\right) ,  \label{eqmax}
\end{equation}%
\begin{equation}
D^{\nu }\rho _{\nu \mu }-m^{2}\rho _{\mu }+iq\gamma \rho ^{\nu }F_{\nu \mu
}=0,  \label{eqvector}
\end{equation}%
The boundary value of $\rho _{\mu }$ is the origin of a charged
vector operator which its expectation value plays the role of
order parameter in the boundary theory. When the temperature
decreases below the critical value, the normal phase becomes
unstable and the vector hair which corresponds to superconducting
phase appears.

In order to study the one-dimensional holographic $p$-wave
superconductor in the presence of backreaction, we take the
following metric for the background geometry,
\begin{equation}
{ds}^{2}=-f(r)e^{-\chi (r)}{dt}^{2}+\frac{{dr}^{2}}{f(r)}+r^{2}{dx}^{2},
\label{metric}
\end{equation}%
with the following choices for the vector and gauge fields,
\begin{equation}\label{rhoA}
\rho _{\nu }dx^{\nu }=\rho _{x}(r)dx,\ \ \ A_{\nu }dx^{\nu }=\phi (r)dt.
\end{equation}%
The Hawking temperature of this black hole is given by \cite{mahya}
\begin{equation}
T=\frac{e^{-\chi (r_{+})/2}f^{^{\prime }}(r_{+})}{4\pi }.  \label{temp}
\end{equation}%
Substituting metric (\ref{metric}) and relation (\ref{rhoA}) in
the field equations (\ref{Eein}) and (\ref{eqmax}), we arrive at
\begin{equation}
\phi ^{\prime \prime }(r)+\left[ \frac{1}{2}\chi ^{\prime }(r)+\frac{1}{r}%
\right] \phi ^{\prime }(r)-\frac{2q^{2}\text{$\rho _{x}$}(r)^{2}\phi (r)}{%
r^{2}f(r)}=0,  \label{phir}
\end{equation}%
\begin{equation}
\rho^{\prime \prime }_{x}(r)+\left[ \frac{f^{\prime }(r)}{f(r)}-%
\frac{1}{2}\chi ^{\prime }(r)-\frac{1}{r}\right] \rho^{\prime }
_{x}(r)+\left[ \frac{q^{2}e^{\chi (r)}\phi (r)^{2}}{f(r)^{2}}-\frac{m^{2}}{f(r)}%
\right] \rho _{x}(r)=0,  \label{rhoxr}
\end{equation}

\begin{equation}
f^{\prime }(r)-\frac{2r}{l^{2}}+\frac{2\kappa ^{2}}{r}\left[ \frac{q^{2}%
\text{$\rho _{x}$}(r)^{2}e^{\chi (r)}\phi (r)^{2}}{f(r)}+f(r)\rho
^{\prime 2}_{x}+m^{2}\rho _{x}(r)^{2}+\frac{r^{2}}{2}e^{\chi
(r)}\phi ^{\prime 2}\right] =0,  \label{fr}
\end{equation}%
\begin{equation}
\chi ^{\prime }(r)+\frac{4\kappa ^{2}}{r}\left[ \frac{q^{2}\text{$\rho _{x}$}%
(r)^{2}e^{\chi (r)}\phi (r)^{2}}{f(r)^{2}}+{\rho^{\prime} _{x}}^2%
\right] =0.  \label{chir}
\end{equation}%
Here, the prime denotes derivative with respect to $r$. If we
consider the probe limit ($\kappa \rightarrow 0$), the equations
of motion (\ref{phir}) and (\ref{rhoxr}) reduce to the
corresponding equations in \cite{alkac}. In the following, we set
$q$ and $l$ equal to unity by using the symmetries
\begin{gather}
q\rightarrow q/a,\text{ \ \ \ \ }\phi \rightarrow a\phi ,\text{ \ \ \ \ }%
\rho _{x}\rightarrow a\rho _{x},\ \ \ \kappa \rightarrow \kappa /a, \\
\notag \\
l\rightarrow al,\text{ \ \ \ \ }r\rightarrow ar,\text{ \ \ \ \ }q\rightarrow
q/a,\ \ \ m\rightarrow m/a.
\end{gather}%
The asymptotic behavior $(r\rightarrow \infty )$ of the solutions
are given by
\begin{equation}
\phi (r)\sim \rho +\mu \ln (r),\ \ f(r)\sim r^{2},\ \ \chi (r)\rightarrow
0,\ \ \rho _{x}(r)\sim \frac{\rho _{x_{-}}}{r^{-m}}+\frac{\rho _{x_{+}}}{%
r^{m}}.  \label{eqasymp}
\end{equation}%
in which $\mu $ and $\rho $ are chemical potential and charge
density, respectively. Note that in (\ref{eqasymp}), the value of
$\chi $\ has been set to zero by virtue of symmetry,
\begin{equation}
e^{\chi }\rightarrow a^{2}e^{\chi },\text{ \ \ \ \ }t\rightarrow at,\text{ \
\ \ \ }\phi \rightarrow \phi /a.  \label{chisym}
\end{equation}%
The asymptotic behavior of the vector field $\rho _{x}(r)$ is in
agreement with the result of \cite{wen18}. Here, the
Breitenlohner-Freedman (BF) bound is $m^{2}\geq 0$. In this limit,
$\rho _{x_{-}}$ plays the role of the source and $\rho _{x_{+}}$
known as $x$-component of the expectation value of the order
parameter $\langle J_{x}\rangle $. In the next sections, we will
analyze the properties of one-dimensional backreacting holographic
$p$-wave superconductor analytically as well as numerically.
\section{Superconductivity phase transition}\label{phase}
In this section, we are going to investigate the phase transition
and critical temperature of ($1+1$)-dimensional backreacting
holographic $p$-wave superconductors. We address the relation
between critical temperature $T_{c}$ and chemical potential $\mu $
as well as the effect of backreaction parameter on $ T_{c}$ in the
vicinity of transition point.
\subsection{Analytical approach}
For the analytical approach, we employ the Sturm-Liouville
eigenvalue problem. To do this we use a coordinate transformation
as $z=r_{+}/r$  where $0\leqslant z\leqslant1$. In the new
coordinates, the field equations (\ref{phir})-(\ref{chir}) turn to
\begin{equation}
\phi ^{\prime \prime }(z)+\left( \frac{\chi ^{\prime }(z)}{2}+\frac{1}{z}%
\right) \phi ^{\prime }(z)-\frac{2\rho _{x}(z)^{2}\phi (z)}{z^{2}f(z)}=0,
\label{phiz}
\end{equation}%
\begin{equation}
\rho _{x}^{\prime \prime }(z)+\left( -\frac{1}{2}\chi ^{\prime }(z)+\frac{%
f^{\prime }(z)}{f(z)}+\frac{3}{z}\right) \rho _{x}^{\prime }(z)+\rho
_{x}(z)\left( \frac{r_{+}^{2}e^{\chi (z)}\phi (z)^{2}}{z^{4}f(z)^{2}}-\frac{%
m^{2}r_{+}^{2}}{z^{4}f(z)}\right) =0,  \label{rhoz}
\end{equation}%
\begin{equation}
f^{\prime }(z)+\frac{2r_{+}^{2}}{l^{2}z^{3}}-2\kappa ^{2}\left( \frac{%
e^{\chi (z)}\phi (z)^{2}\rho _{x}(z){}^{2}}{zf(z)}+\frac{z^{3}f(z)\rho
_{x}^{\prime 2}}{r_{+}^{2}}+\frac{m^{2}\rho _{x}(z){}^{2}}{z}+\frac{z}{2}%
e^{\chi (z)}\phi ^{\prime 2}\right) =0,  \label{fz}
\end{equation}%
\begin{equation}
\chi ^{\prime 2}\left( \frac{e^{\chi (z)}\phi (z)^{2}\rho _{x}(z){}^{2}}{%
zf(z)^{2}}+\frac{z^{3}\rho _{x}^{\prime 2}}{r_{+}^{2}}\right) =0.
\label{chiz}
\end{equation}%
Here, the prime indicates the derivative with respect to $z$. Near
the critical temperature, the expectation value of $\langle
J_{x}\rangle $ is small so we can take it as an expansion
parameter
\begin{equation*}
\epsilon \equiv \left\langle J_{x}\right\rangle.
\end{equation*}%
Since in the vicinity of critical temperature $\epsilon \ll 1$, we
focus on solutions for small values of condensation parameter
$\epsilon $. Therefore,
we can expand the model functions as%
\begin{gather}
\rho _{x}\approx \epsilon \rho _{x_{1}}+\epsilon ^{3}\rho _{x_{3}}+\epsilon
^{5}\rho _{x_{5}}+\cdots , \\
\phi \approx \phi _{0}+\epsilon ^{2}\phi _{2}+\epsilon ^{4}\phi _{4}+\cdots ,
\\
f\approx f_{0}+\epsilon ^{2}f_{2}+\epsilon ^{4}f_{4}+\cdots , \\
\chi \approx \epsilon ^{2}\chi _{2}+\epsilon ^{4}\chi _{4}+\cdots .
\label{chizero}
\end{gather}%
Furthermore, we have a similar expression for the chemical
potential which can be expressed as
\begin{equation}
\mu =\mu _{0}+\epsilon ^{2}\delta \mu _{2}+...\rightarrow \epsilon
\thickapprox \Bigg(\frac{\mu -\mu _{0}}{\delta \mu _{2}}\Bigg)^{1/2},
\end{equation}%
where $\delta \mu _{2}>0$. Thus near the phase transition point
($\mu _{c}=\mu _{0}$), the order parameter $\epsilon$ vanishes. In
addition, we obtain the mean field value of the critical exponent
as $\beta =1/2$.

The equation of motion for the gauge field (\ref{phiz}) at zeroth
order of $\epsilon $ is given by
\begin{equation}
\phi ^{\prime \prime }(z)+\frac{\phi ^{\prime }(z)}{z}=0.  \label{phiii}
\end{equation}%
The solutions of this equation reads
\begin{equation}
\phi (z)=\lambda r_{+}\log (z),\ \ \ \lambda =\frac{\mu }{r_{+}}.
\label{phiii2}
\end{equation}%
Combining the solutions (\ref{phiii2}) with Eq. (\ref{fz}), the equation  for $%
f(z)$, at zeroth order of $\epsilon $, can be obtained as%
\begin{equation}
f^{\prime }(z)+\frac{2r_{+}^{2}}{z^{3}}-\frac{\kappa ^{2}\lambda
^{2}r_{+}^{2}}{z}=0,  \label{ffz}
\end{equation}%
which has the solutions,
\begin{equation}
f(z)=\frac{r_{+}^{2}g(z)}{z^{2}},\text{ \ \ \ \ }g(z)=1-z^{2}+\kappa
^{2}\lambda ^{2}z^{2}\log (z).  \label{eqgz}
\end{equation}%
Near the boundary, the vector field can be defined as:
\begin{equation}
\rho _{x}(z)=\frac{\langle J_{x}\rangle }{\sqrt{2}r_{+}^{\Delta }}z^{\Delta
}F(z).  \label{eqrhoo}
\end{equation}%
Inserting Eqs. (\ref{eqgz}) and (\ref{eqrhoo}) in Eq. (\ref{rhoz})
yields to
\begin{equation}
F^{\prime \prime }(z)+F^{\prime }(z)\left( \frac{g^{\prime }(z)}{g(z)}+\frac{%
2\Delta }{z}+\frac{1}{z}\right) +F(z)\left( \frac{\Delta g^{\prime }(z)}{%
zg(z)}-\frac{m^{2}}{z^{2}g(z)}+\frac{\Delta ^{2}}{z^{2}}\right) +\frac{%
F(z)\left( \lambda ^{2}\kappa ^{2}\log ^{2}(z)\right) }{g(z)^{2}}=0.
\label{eqF}
\end{equation}%
If we define some new functions as below, we can rewrite Eq.
(\ref{eqF}) in the Sturm-Liouville form as
\begin{equation}
\left[ T(z)F^{\prime }(z)\right] ^{\prime }+P(z)T(z)F(z)+\lambda
^{2}Q(z)T(z)F(z)=0,  \label{sl}
\end{equation}%
where
\begin{equation}
T(z)=z^{2\Delta +1}g(z),\text{ \ \ \ \ }P(z)=\left[ \frac{\Delta g^{\prime
}(z)}{zg(z)}-\frac{m^{2}}{z^{2}g(z)}+\frac{\Delta ^{2}}{z^{2}}\right] ,\text{
\ \ \ \ }Q(z)=\frac{\log ^{2}(z)}{g(z)^{2}}.
\end{equation}%
Next, we define a trial function $F(z)=1-\alpha z^{2}$ which is
satisfied in the boundary conditions $F(0)=1$ and $F^{^{\prime
}}(0)=0$. By minimizing the following expression with respect to
$\alpha $, equation (\ref{sl}) will be solved:
\begin{equation}
\lambda ^{2}=\frac{\int_{0}^{1}T\left( F^{\prime 2}-PF^{2}\right) dz}{%
\int_{0}^{1}TQF^{2}dz},  \label{l2}
\end{equation}%
With the help of iteration method, definition of backreacting
parameter is \cite{LPJW2015}
\begin{equation}
\kappa _{n}=n\Delta \kappa ,\ \ \ n=0,1,2,\cdots ,\ \ \Delta \kappa =\kappa
_{n+1}-\kappa _{n}.
\end{equation}%
Here, $\Delta \kappa =0.05$. In addition, we have%
\begin{equation}
\kappa ^{2}\lambda ^{2}={\kappa _{n}}^{2}\lambda ^{2}={\kappa _{n}}%
^{2}(\lambda ^{2}|_{\kappa _{n-1}})+O[(\Delta \kappa )^{4}],
\end{equation}%
where $\kappa _{-1}=0$ and $\lambda ^{2}|_{\kappa _{-1}}=0$. At the critical
point, at zeroth order with respect to $\epsilon $, critical temperature is
defined as\footnote{%
Note that $\chi $ tends to zero near the critical point according to (\ref%
{chizero}).}
\begin{equation}
T_{c}=\frac{f^{\prime }\left( r_{+c}\right) }{4\pi }=r_{+c}\left( \frac{2-{%
\kappa }^{2}\lambda ^{2}}{4\pi }\right) =\frac{\mu }{\lambda }\left( \frac{2-%
{\kappa }_{n}^{2}\lambda ^{2}|_{\kappa _{n-1}}}{4\pi }\right) .  \label{tc}
\end{equation}%
The analytical results of $T_{c}/\mu $ for different values of
mass and backreaction parameters are shown in table \ref{tab1}.
According to these results, enlarging the values of mass have the
same effect as increasing backreaction parameter on $T_{c}/\mu $
and makes it smaller. So, it causes condensation harder to form.
\subsection{Numerical Method}
We employ the shooting method \cite{H09} to numerically
investigate the properties of ($1+1$)-dimensional holographic
$p$-wave superconductor developed in BTZ black hole background,
when the gauge and vector fields backreact on the background
geometry. For this purpose, we must know the behavior of the model
functions both at horizon and boundary. By using Taylor expansion
around horizon we arrive at

\begin{gather}
f(z)=f_{1}\left( 1-z\right) +f_{2}\left( 1-z\right) {}^{2}+\cdots , \\
\phi (z)=\phi _{1}\left( 1-z\right) +\phi _{2}\left( 1-z\right)
{}^{2}+\cdots , \\
\rho _{x}(z)=\rho _{x_{0}}+\rho _{x_{1}}\left( 1-z\right) +\rho
_{x_{2}}\left( 1-z\right) {}^{2}+\cdots , \\
\chi (z)=\chi _{0}+\chi _{1}\left( 1-z\right) +\chi _{2}\left( 1-z\right)
{}^{2}+\cdots .
\end{gather}%
We impose the boundary condition $\phi (z=1)$ which is motivated
from the fact that the gauge field $A_{\nu}$ has a finite
norm at the horizon. In this method, all coefficients will be defined in terms of $\phi _{1}$, $%
\rho _{x_{0}}$ and $\chi _{0}$. Our desirable state is $\rho
_{x_{-}}(\infty )=\chi (\infty )=0$. This will be achieved by
varying $\phi _{1}$, $\rho _{x_{0}}$ and $\chi _{0}$ at the
horizon. Furthermore, we can set $r_{+}=1$ by virtue
of the equations of motion's symmetry%
\begin{equation}
r\rightarrow ar,\text{ \ \ \ \ }f\rightarrow a^{2}f,\text{ \ \ \ \ }\phi
\rightarrow a\phi .
\end{equation}%
\begin{figure*}[t]
\centering
\subfigure[~$m^{2}=\frac{1}{16}$]{\includegraphics[width=0.3%
\textwidth]{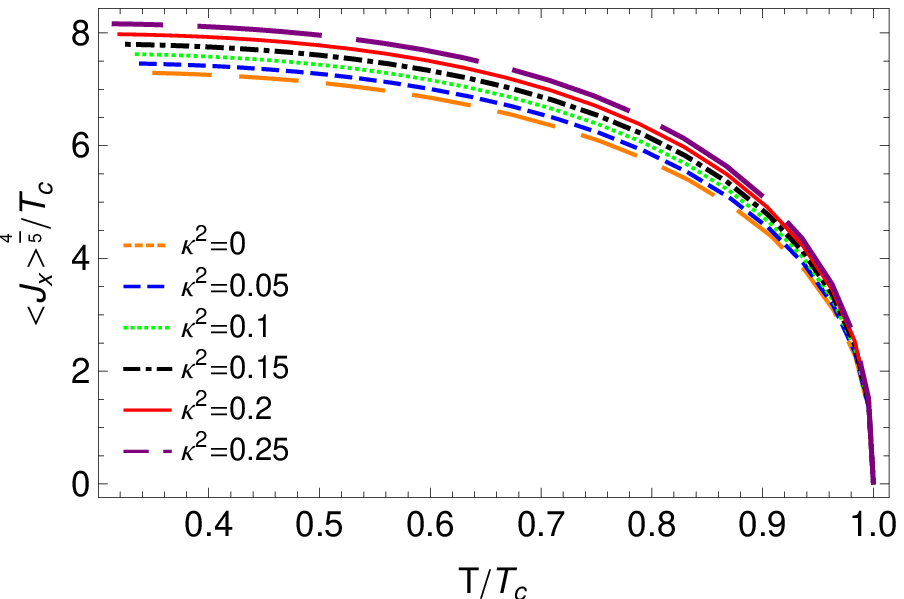}} \qquad \subfigure[~$m^{2}=\frac{1}{4}$]{%
\includegraphics[width=0.3\textwidth]{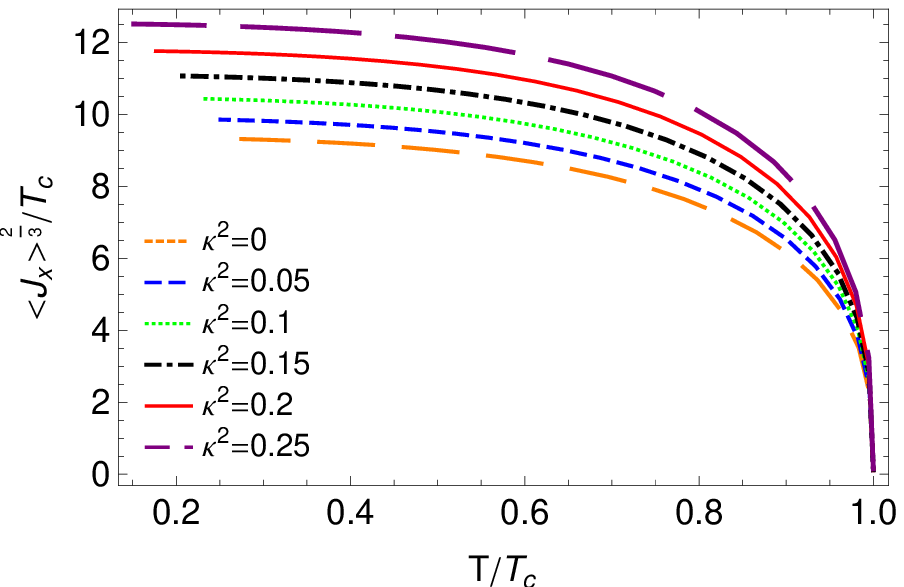}} \qquad %
\subfigure[~$m^{2}=1$]{\includegraphics[width=0.3\textwidth]{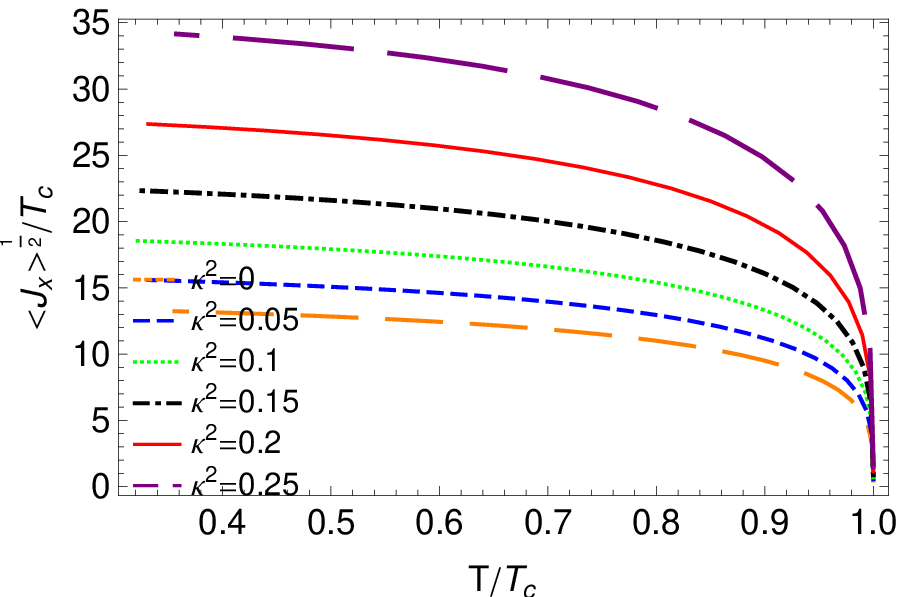}}
\caption{The behavior of condensation parameter as a function of temperature
for different values of backreaction.}
\label{fig1}
\end{figure*}
The consequence of this method is finding the values of $T_{c}/\mu
$ for different masses and backreaction parameters. In order to
compare the numerical and analytical results, data are given in
table \ref{tab1}. The results of Sturm-Liouville method are
confirmed by numerical data. The effects of mass and backreaction
parameters on the behaviour of condensation are shown in Fig.
\ref{fig1}. We see that all curves follow the same behaviour. As
it is clear from Fig. \ref{fig1}, enhancing the values of mass and
backreaction parameter causes the gap in curves larger and so it
makes the formation of condensation harder. As a result, the
critical temperature decreases with increasing the backreaction
and mass parameters.
\begin{table*}[t]
\begin{center}
\begin{tabular}{c|c|c|c|c|c|c|}
\cline{2-3}\cline{2-7}\cline{4-7}
& \multicolumn{2}{|c|}{$m^{2}=\dfrac{1}{16}$} & \multicolumn{2}{|c|}{$m^{2}=%
\dfrac{1}{4}$} & \multicolumn{2}{|c|}{$m^{2}=1$} \\
\cline{2-3}\cline{2-7}\cline{4-7}
& Analytical & Numerical & Analytical & Numerical & Analytical & Numerical
\\ \hline
\multicolumn{1}{|c|}{$\kappa ^{2}=0$} & $0.1424$ & $0.1433 $ & $0.0860$ & $%
0.0880$ & $0.0478$ & $0.0503$ \\ \hline
\multicolumn{1}{|c|}{$\kappa ^{2}=0.05$} & $0.1363$ & $0.1397$ & $0.0814$ & $%
0.0823$ & $0.0443$ & $0.0410$ \\ \hline
\multicolumn{1}{|c|}{$\kappa ^{2}=0.1$} & $0.1356$ & $0.1361$ & $0.0803$ & $%
0.0768$ & $0.0424$ & $0.0330$ \\ \hline
\multicolumn{1}{|c|}{$\kappa ^{2}=0.15$} & $0.1346$ & $0.1326 $ & $0.0786$ &
$0.0716$ & $0.0394$ & $0.0260$ \\ \hline
\multicolumn{1}{|c|}{$\kappa ^{2}=0.2$} & $0.1332$ & $0.1292$ & $0.0763$ & $%
0.0666$ & $0.0353$ & $0.0201$ \\ \hline
\multicolumn{1}{|c|}{$\kappa ^{2}=0.25$} & $0.1313$ & $0.1258$ & $0.0733$ & $%
0.0619$ & $0.0302$ & $0.0152$ \\ \hline
\end{tabular}%
\end{center}
\caption{Analytical and Numerical results of ${T_{c}}/{\protect\mu
}$ for different values of backreaction and mass
parameters.}\label{tab1}
\end{table*}
\section{Critical exponents\label{Crit}}
In this section we are going to calculate the expectation value of
$\langle J_{x}\rangle $ in the boundary theory near the critical
temperature for one-dimensional  holographic $p$-wave
superconductor in the presence of backreaction. Furthermore, we
compute the values of the critical exponents both analytically and
numerically.
\subsection{Analytical approach}
We focus on the behavior of the gauge field in the vicinity of the
critical temperature. In this limit, the field equation
(\ref{phiz}) turns to
\begin{equation}
\phi ^{\prime \prime }(z)+\frac{\phi ^{\prime }(z)}{z}-\frac{2\rho
_{x}(z)^{2}\phi (z)}{z^{2}f(z)}=0.  \label{phii}
\end{equation}%
Because of nonzero value of the condensation in the vicinity of
the critical temperature, we have an extra term in the above
equation in comparison with the field equation in the previous
section. Inserting Eqs. (\ref{eqgz}) and (\ref{eqrhoo}) in Eq.
(\ref{phii}) we have
\begin{equation}
\phi ^{\prime \prime }(z)+\frac{\phi ^{\prime }(z)}{z}=\frac{\langle
J_{x}\rangle ^{2}z^{2\Delta }}{r_{+}^{2\Delta +2}g(z)}F(z)^{2}\phi (z),
\label{phi2}
\end{equation}%
Using the fact that the value of $\frac{\langle J_{x}\rangle ^{2}}{%
r_{+}^{2\Delta +2}}$ is small in $T\sim T_{c}$ limit, we assume
that Eq. (\ref{phi2}) has the following answer
\begin{equation}
\frac{\phi (z)}{r_{+}}=\lambda \log (z)+\frac{\langle J_{x}\rangle ^{2}}{%
r_{+}^{2\Delta +2}}\eta (z),\ \ \ \lambda =\frac{\mu }{r_{+}}.  \label{phi3}
\end{equation}%
Since at the horizon $\phi(z=1)=0$, thus we have $\eta (1)=0$.
Substituting the above
equation in Eq. (\ref{phi2}) up to $\frac{\langle J_{x}\rangle ^{2}}{%
r_{+}^{2\Delta +2}}$ order, we arrive at
\begin{equation}
\eta ^{\prime \prime }(z)+\frac{\eta ^{\prime }(z)}{z}=\frac{\lambda
z^{2\Delta }\log (z)}{g(z)}F(z)^{2}.  \label{eta}
\end{equation}%
Multiplying the both sides of Eq. (\ref{eta}) by $z$ and
integrating from $z=0$ to $z=1$, we get
\begin{equation}
\int_{0}^{1}d\left( z\eta ^{\prime }(z)\right) =\eta ^{\prime
}(1)=\lambda \int_{0}^{1}\frac{z^{2\Delta +1}\log
(z)}{1-z^{2}+\kappa ^{2}\lambda ^{2}z^{2}\log
(z)}F(z)^{2}\,dz=\lambda \mathcal{A},  \label{etap}
\end{equation}%
where
\begin{equation}
\mathcal{A}\equiv\int_{0}^{1}\frac{z^{2\Delta +1}\log
(z)}{1-z^{2}+\kappa ^{2}\lambda ^{2}z^{2}\log (z)}F(z)^{2}\,dz.
\label{A}
\end{equation}%
Combining Eqs. (\ref{eqasymp}) and (\ref{phi3}) and taking into
account the fact that the first term on the rhs of Eq.
(\ref{phi3}) is the solution of $\phi (z)$ at the critical point,
and the second term is a correction term, we can write near the
critical point,
\begin{equation}
\frac{\rho }{r_{+}}+\frac{\mu }{r_{+}}\log (z)=+\frac{\mu }{r_{+c}}\log (z)+%
\frac{\langle J_{x}\rangle ^{2}}{r_{+c}^{2\Delta +2}}\eta (z),  \label{eqq0}
\end{equation}%
Now, we use a coordinate transformation $%
z\rightarrow Z+1$, then by expanding the resulting equation around $Z=0$ we
get
\begin{equation}
\frac{\rho }{r_{+}}+\frac{\mu }{r_{+}}\left(Z-\frac{Z^{2}}{2}+...\right)=+\frac{\mu }{%
r_{+c}}\left(Z-\frac{Z^{2}}{2}+...\right)+\frac{\langle
J_{x}\rangle ^{2}}{r_{+c}^{2\Delta +2}}\left(\eta (1)+Z\eta
^{\prime }(1)+...\right). \label{eqq}
\end{equation}%
Comparing the coefficients $Z$ on both sides of Eq. (\ref{eqq})
and using Eq. (\ref{etap}) we find
\begin{equation}
\frac{\mu }{r_{+}}=\frac{\mu }{r_{+c}}\left( 1+\frac{\langle J_{x}\rangle
^{2}}{r_{+c}^{2\Delta +2}}\mathcal{A}\right) ,  \label{samez}
\end{equation}%
Near the critical point we have $T\sim T_{c}$, and thus using
relation (\ref{tc}), we can find the equation of $r_{+}$ as below
\begin{equation}
r_{+}=\frac{4\pi T}{\left( 2-\kappa ^{2}\lambda ^{2}\right) },  \label{eqr+}
\end{equation}%
Inserting Eqs. (\ref{tc}) and (\ref{eqr+}) in Eq. (\ref{samez})
and taking the absolute values of the resulting equation, we
arrive at
\begin{equation}
\langle J_{x}\rangle =\gamma T_{c}^{\Delta
+1}\sqrt{1-\frac{T}{T_{c}}}, \label{eqJx}
\end{equation}%
where
\begin{equation}
\gamma =\frac{1}{\sqrt{\left\vert \mathcal{A}\right\vert }}\left(
\frac{4\pi }{2-\kappa ^{2}\lambda ^{2}}\right) ^{\Delta +1}.
\end{equation}
Based on the above equation, it is obvious that the critical
exponent $\beta =1/2$ is in a perfect agreement with the mean
field theory. Since the value of $\beta $ is independent of the
effect of backreaction, we have the second order phase transition
for all values of the backreaction parameter. The analytical
results are shown in table \ref{table2}. Increasing the values of
the mass and backreaction parameters, causes the larger values of
the condensation parameter. Therefore, the larger values of the
mass as well as the backreacting of the gauge and vector fields on
the background geometry makes the condensation harder to form.
\subsection{Numerical Method}
\begin{figure*}[t]
\centering
\subfigure[~$m^{2}=\frac{1}{16}$]{\includegraphics[width=0.3%
\textwidth]{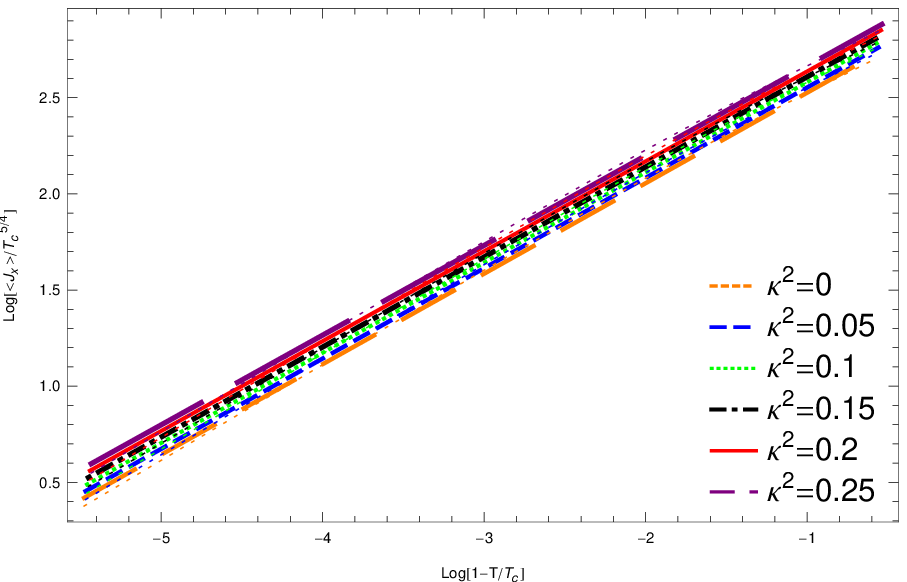}} \qquad \subfigure[~$m^{2}=\frac{1}{4}$]{%
\includegraphics[width=0.3\textwidth]{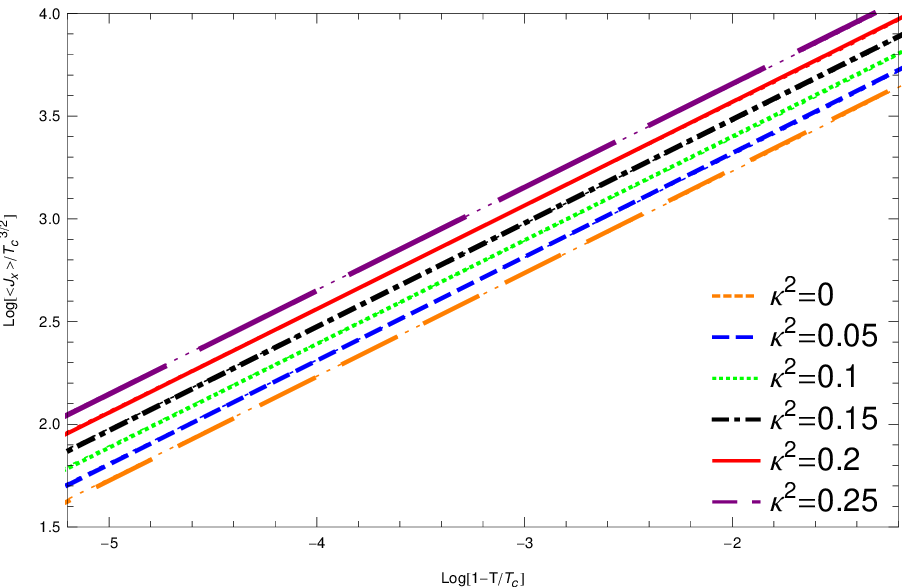}} \qquad %
\subfigure[~$m^{2}=1$]{\includegraphics[width=0.3\textwidth]{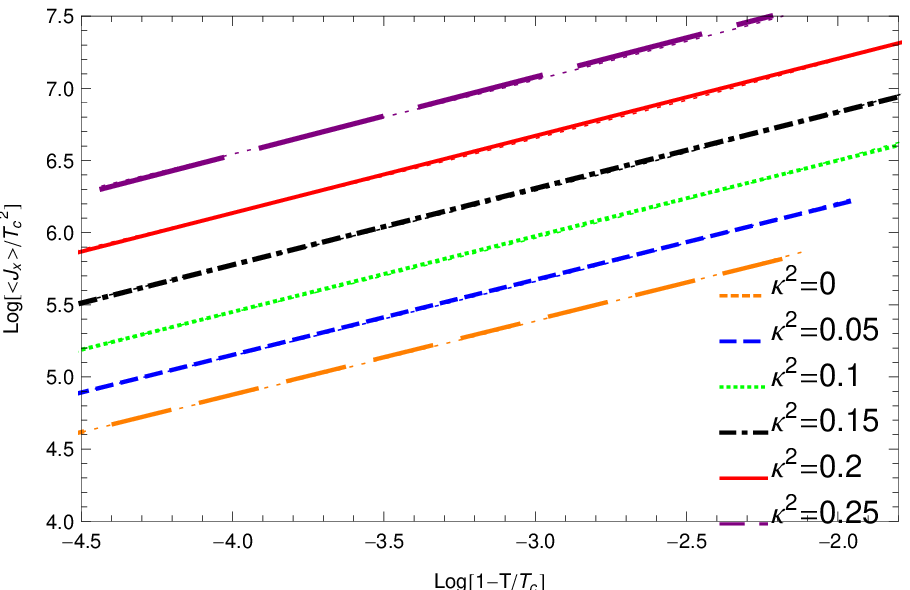}}
\caption{The behavior of $\log \langle J_{x}\rangle /T_{c}^{\Delta
+1}$ as a function of $\log \left( 1-T/T_{c}\right) $ with slope
of $1/2$ for different values of mass and backreaction
parameters.} \label{fig2}
\end{figure*}

Based on the behavior of condensation near the critical
temperature which is obtained by using analytical approach (i.e.
Eq. (\ref{eqJx})) we have
\begin{equation}
\log \left( \frac{\langle J_{x}\rangle }{T_{c}^{\Delta +1}}\right) =\log
(\gamma )+\frac{1}{2}\log \left( 1-\frac{T}{T_{c}}\right) .  \label{eqlogjx}
\end{equation}%
In Fig \ref{fig2}, the behavior of $\log \langle J_{x}\rangle
/T_{c}^{\Delta +1}$ as a function of $\log \left( 1-T/T_{c}\right)
$ for different values of the backreaction and mass parameters was
shown. The slope of curves is $1/2$ which is in agreement with
mean field theory and shows that we face with a second order phase
transition as same as the analytical approach. In addition, this
value of critical exponent is independent of the backreaction
parameter. Using Eq. (\ref{eqlogjx}), it is obvious that the
intercept of curves represent the values of $\log \gamma $,
numerically. In order to compare the analytical and numerical
values of $\gamma $, the results are listed in table \ref{table2}.
The most agreement between the values of $\gamma $ from these two
approaches appears in $m^{2}=1/16$ and for larger values of mass
we observe less match. In addition, the values of $\gamma $
increase for larger values of backreaction parameter. Same results
are obtained in analytical method, too.

\begin{table*}[t]
\begin{center}
\begin{tabular}{c|c|c|c|c|c|c|}
\cline{2-3}\cline{2-7}\cline{4-7}
& \multicolumn{2}{|c|}{$m^{2}=\dfrac{1}{16}$} & \multicolumn{2}{|c|}{$m^{2}=%
\dfrac{1}{4}$} & \multicolumn{2}{|c|}{$m^{2}=1$} \\
\cline{2-3}\cline{2-7}\cline{4-7}
& Analytical & Numerical & Analytical & Numerical & Analytical & Numerical
\\ \hline
\multicolumn{1}{|c|}{$\kappa ^{2}=0$} & $23.5779$ & $20.0224 $ & $48.1780$ &
$70.3273$ & $188.5$ & $1042.6100$ \\ \hline
\multicolumn{1}{|c|}{$\kappa ^{2}=0.05$} & $23.7791$ & $20.5590$ & $49.1955$
& $76.0058$ & $199.517$ & $1390.0400$ \\ \hline
\multicolumn{1}{|c|}{$\kappa ^{2}=0.1$} & $23.9228$ & $21.1162$ & $50.2391$
& $82.3205$ & $220.382$ & $1901.3500$ \\ \hline
\multicolumn{1}{|c|}{$\kappa ^{2}=0.15$} & $24.1599$ & $21.6954 $ & $51.9956$
& $89.3687$ & $260.302$ & $2682.0100$ \\ \hline
\multicolumn{1}{|c|}{$\kappa ^{2}=0.2$} & $24.4981$ & $22.2974$ & $54.6113$
& $97.2622$ & $333.956$ & $3928.0200$ \\ \hline
\multicolumn{1}{|c|}{$\kappa ^{2}=0.25$} & $24.9451$ & $22.9237$ & $58.2727$
& $106.138$ & $475.166$ & $6028.6000$ \\ \hline
\end{tabular}%
\end{center}
\caption{Analytical and Numerical results of $\protect\gamma $ for
different values of backreaction and mass parameters.}
\label{table2}
\end{table*}
\section{Closing remarks}
In this paper, we analyzed a holographic $p$-wave superconductor
model in a three-dimensional Einstein-Maxwell theory in the
presence of negative cosmological constant and a vector field when
the gauge and vector fields backreact on the background geometry.
In order to study the problem analytically, we employ the
Sturm-Liouville eigenvalue problem while the numerical data were
achieved with help of shooting method. We analytically calculated
the relation between the critical temperature and chemical
potential for different values of the mass and backreaction
parameters. These data were confirmed by numerical results. We
found out that increasing the values of the mass and backreaction
parameters makes the condensation harder to form and thus the
critical temperature decreases. In addition, critical exponent of
this system have also obtained both analytically and numerically.
Based on these investigations, it was expressed that we face a
second order phase transition. Furthermore, the obtained critical
exponent value $\beta ={1}/{2}$ follows the mean field theory
value. Since the nonlinear electrodynamics give more information
in comparison with Maxwell case, it is worthwhile to consider the
effect of nonlinearity on the physical properties of holographic
$p$-wave superconductors. We leave this issue for future
investigations.
\begin{acknowledgments}
We thank Shiraz University Research Council. The work of AS has
been supported financially by Research Institute for Astronomy and
Astrophysics of Maragha (RIAAM), Iran. MKZ thanks Shahid Chamran
University of Ahvaz for supporting this work.
\end{acknowledgments}


\begin{thebibliography}{Horowitz et al.(2008)}
\bibitem{Dahi}  Per F. Dahl, Historical Studies in the Physical Sciences, Vol. {\bf15},
No. 1 (1984) 1.

\bibitem{alkac} G. Alkac, S. Chakrabortty, P. Chaturvedi, Phys. Rev. D
\textbf{96}, 086001 (2017) [arXiv:1610.08757].

\bibitem{BCS57} J. Bardeen, L. N. Cooper, J. R. Schrieer, Phys. Rev. \textbf{%
108}, 1175 (1957). 

\bibitem[Maldacena.(1998)]{Maldacena} J. M. Maldacena, Adv. Theor. Math.
Phys. \textbf{2}, 231 (1998) [hep-th/9711200v3]

\bibitem{H08} S. A. Hartnoll, C. P. Herzog and G. T. Horowitz, Phys. Rev.
Lett. \textbf{101}, 031601 (2008) [arXiv:0803.3295].


\bibitem[Gubser et al.(1998)]{G98} S. S. Gubser, I. R. Klebanov and A. M.
Polyakov, Phys. Lett. B \textbf{428}, 105 (1998) [hep-th/9802109].


\bibitem[Witten et al.(1998)]{W98} E. Witten, Adv. Theor. Math. Phys.
\textbf{2}, 253 (1998) [hep-th/9802150].




\bibitem[Horowitz et al.(2008)]{HR08} G. T. Horowitz and M. M. Roberts,
Phys. Rev. D \textbf{78}, 126008 (2008).

\bibitem[Ren.(2010)]{R10} J. Ren, JHEP. \textbf{1011}, 055 (2010)
[arXiv:1008.3904].

\bibitem[Hartnoll(2009)]{H09} S. A. Hartnoll, Class. Quantum Grav. \textbf{26%
}, 224002 (2009) [arXiv:0903.3246].


\bibitem[Herzog.(2009)]{Hg09} C. P. Herzog, J. Phys. A \textbf{42}, 343001
(2009) [arXiv:0904.1975].


\bibitem[Horowitz(2011)]{H11} G. T. Horowitz, Lect. Notes Phys. \textbf{828}%
, 313 (2011) [arXiv:1002.1722].

\bibitem[Gubser.(2009)]{Gu09} S. S. Gubser, C. P. Herzog, S. S. Pufu and T.
Tesileanu, Phys. Rev. Lett. \textbf{103}, 141601 (2009) [arXiv:0907.3510].

\bibitem[HHH.(2008)]{HHH08} S. A. Hartnoll, C. P. Herzog and G. T. Horowitz,
JHEP \textbf{0812}, 015 (2008) [arXiv:0810.1563].

\bibitem[Jing ,Chena(210)]{JCH10} J. Jing, S. Chen, Phys. Lett. B \textbf{686%
}, 68 (2010) [arXiv:1001.4227].

\bibitem[cai(2015)]{cai15} R. G. Cai, L. Li, Li-Fang Li, Run-Qiu Yang, Sci
China Phys. Mech. Astron. \textbf{58}, 060401 (2015) [arXiv:1502.00437].

\bibitem[Ge(2010)]{Ge10} X. H. Ge, B. Wang, S. F. Wu, and G. H. Yang, JHEP
\textbf{1008}, 108 (2010) [arXiv:1002.4901].

\bibitem[Ge(2012)]{Ge12} X. H. Ge, S. F. Tu, B. Wang, JHEP \textbf{09}, 088
(2012) [arXiv:1209.4272].

\bibitem[Kuang(2013)]{Kuang13} X. M. Kuang, E. Papantonopoulos, G. Siopsis,
B. Wang, Phys. Rev. D \textbf{88}, 086008 (2013) [arXiv:1303.2575].

\bibitem[Pan(2011)]{Pan11} Q. Pan, J. Jing, B. Wang, JHEP \textbf{11}, 088
(2011) [arXiv:1105.6153].

\bibitem{Wang6} M. Kord Zangeneh, Y. C. Ong, B. Wang, Phys. Lett. B \textbf{%
771}, 235 (2017) [arXiv:1704.00557].

\bibitem[CAI(11)]{CAI11} R. G. Cai, H. F Li, H.Q. Zhang, Phys. Rev. D
\textbf{83}, 126007 (2011).

\bibitem[cai(10)]{cai10} R. G. Cai, Z.Y. Nie, H.Q. Zhang, Phys. Rev. D
\textbf{82}, 066007 (2010).


\bibitem[yao(2013)]{yao13} W. Yao, J. Jing, JHEP \textbf{1305}, 101 (2013)
[arXiv:1306.0064].

\bibitem{n4} Z. Zhao, Q. Pan, S. Chen and J. Jing, Nucl. Phys. B \textbf{871}%
, 98 (2013) [arXiv:1212.6693].

\bibitem{n6} Y. Liu, Y. Gong and B. Wang, JHEP \textbf{1602}, 116 (2016)
[arXiv:1505.03603].

\bibitem{Gan1} S. Gangopadhyay, D. Roychowdhury, JHEP \textbf{05}, 002
(2012) [arXiv:1201.6520];\newline
S. Gangopadhyay and D. Roychowdhury, JHEP \textbf{05}, 156 (2012)
[arXiv:1204.0673].

\bibitem[Sheykhi et al.(2016)]{SH16} A. Sheykhi, H. R. Salahi, A. Montakhab,
JHEP \textbf{1604}, 058 (2016) [arXiv:1603.00075].

\bibitem[Salahi et al.(2016)]{SSh16} H. R. Salahi, A. Sheykhi, A. Montakhab,
Eur. Phys. J. C \textbf{76}, 575 (2016) [arXiv:1608.05025].

\bibitem[SHsh(17)]{SHsh(17)} A. Sheykhi, F. Shaker, Int. J.
Mod. Phys. D \textbf{26}, 1750050 (2017) [arXiv:1606.04364].

\bibitem[SHSH(16)]{SHSH(16)} A. Sheykhi, F. Shaker, Can. J. of Phys.
\textbf{94}, 1372 (2016) [arXiv:1601.05817].

\bibitem{shSh(16)} A. Sheykhi, F. Shaker, Phys Lett. B \textbf{754}, 281
(2016) [arXiv:1601.04035].

\bibitem{Doa} A. Sheykhi, D. Hashemi Asl, A. Dehyadegari, Phys. Lett. B
\textbf{781}, 139 (2018) [arXiv:1803.05724].

\bibitem{Afsoon} A. Sheykhi, A. Ghazanfari, A. Dehyadegari, Eur. Phys. J. C
\textbf{78}, 159 (2018) [arXiv:1712.04331].

\bibitem{n3} M.~Kord Zangeneh, S.~S.~Hashemi, A.~Dehyadegari, A.~Sheykhi and
B.~Wang, Phys. Lett. B \textbf{785}, 238 (2018) [arXiv:1710.10162].

\bibitem{n5} S. I. Kruglov, arXiv:1801.06905.

\bibitem{Gubser} S. S. Gubser and S. S. Pufu,  JHEP, {\bf0811}, 033 (2008).

\bibitem{Donos} A. Donos and J.P. Gauntlett, JHEP {\bf12}, 091 (2011).

\bibitem{Caip} R.-G. Cai, S. He, L. Li, and L.F. Li.  JHEP,
{\bf1312}, 036 (2013).

\bibitem{cai13p} R.G.Cai, L. Li, L.F. Li, JHEP, \textbf{1401}, 032 (2014)
[arXiv:1309.4877v3]. 

\bibitem{Roberts8} M. M. Roberts and S. A. Hartnoll, JHEP \textbf{0808}, 035
(2008) [arXiv:0805.3898].

\bibitem{zeng11} H. B. Zeng, W. M. Sun and H. S. Zong, Phys. Rev. D \textbf{%
83}, 046010 (2011) [arXiv:1010.5039 [hep-th]].

\bibitem{cai11p} R. G. Cai, Z. Y. Nie and H. Q. Zhang, Phys. Rev. D \textbf{%
83}, 066013 (2011) [arXiv:1012.5559].

\bibitem{pando12} L. A. Pando Zayas and D. Reichmann, Phys. Rev. D \textbf{85%
}, 106012 (2012) [arXiv:1108.4022].

\bibitem{momeni12p} D. Momeni, N. Majd and R. Myrzakulov, Europhys. Lett.
\textbf{97}, 61001 (2012) [arXiv:1204.1246].



\bibitem{gangopadhyay12} S. Gangopadhyay, D. Roychowdhury, JHEP \textbf{08},
104 (2012) [arXiv:1207.6505v2].

\bibitem{chaturverdip15} P. Chaturvedi, G. Sengupta, JHEP, \textbf{1504},
001 (2015) [arXiv:1501.06998v1].

\bibitem{Car1} S. Carlip, Class. Quant. Grav. \textbf{12}, 2853 (1995)
[gr-qc/9506079]. 

\bibitem{Ash} A. Ashtekar, J. Wisniewski and O. Dreyer, Adv. Theor. Math.
Phys. \textbf{6}, 507 (2002) [gr-qc/0206024].

\bibitem{Sar} T. Sarkar, G. Sengupta and B. Nath Tiwari, JHEP \textbf{0611},
015 (2006) [hep-th/0606084].

\bibitem{Wit1} E. Witten, Adv. Theor. Math. Phys. \textbf{2}, 505 (1998)
[hep-th/9803131].

\bibitem{Car2} S. Carlip, Class. Quant. Grav. \textbf{22}, R85 (2005)
[gr-qc/0503022].

\bibitem{Bu} Y. Bu,  Phys. Rev. D {\bf86}, 106005 (2012).



\bibitem{Wit2} E. Witten, arXiv:0706.3359.

\bibitem{chaturvedi} P. Chaturvedi, G. Sengupta, Phys. Rev. D \textbf{90},
046002 (2014) [arXiv:1310.5128].

\bibitem[Li(2012)]{L12} R. Li, Mod. Phys. Lett. A. \textbf{27}, 1250001
(2012).

\bibitem{momeni} D. Momeni, M. Raza, M. R. Setare, R. Myrzakulov, Int. J.
Theor. Phys. \textbf{52}, 2773 (2013) [arXiv:1305.5163].

\bibitem{peng17} Y. Peng, G. liu, Int. J. Mod. Phys. A \textbf{32}, 1750160
(2017).

\bibitem{Wang} Y. Liu, Q. Pan and B. Wang, Phys. Lett. B \textbf{702}, 94
(2011) [arXiv:1106.4353].

\bibitem{lashkari} N. Lashkari, JHEP \textbf{1111}, 104 (2011)
[arXiv:1011.3520].

\bibitem{hua} H. B. Zeng, arXiv:1204.5325.

\bibitem{yanyan} Y. Bu, Phys. Rev. D \textbf{86}, 106005 (2012)
[arXiv:1205.1614].

\bibitem{yan} Y. Peng, [arXiv:1604.06990].

\bibitem{kord} M. Kord Zangeneh, Y. C. Ong and B. Wang, Phys. Lett. B
\textbf{771}, 235 (2017) [arXiv:1704.00557].

\bibitem{bina} B. Binaei Ghotbabadi, M. Kord Zangeneh and A. Sheykhi, Eur.
Phys. J. C \textbf{78}, 381 (2018) [arXiv:1804.05442].

\bibitem{mahya} M. Mohammadi, A. Sheykhi and M. Kord Zangeneh, Eur. Phys. J.
C \textbf{78}, 654 (2018) [arXiv:1805.07377v1].

\bibitem{wen18} D. Wen, H. Yu, Q. Pan, K. Lin and W. L. Qian, Nucl. Phys. B
\textbf{930}, 255 (2018) [arXiv:1803.06942v2].

\bibitem{LPJW2015} C. Lai, Q. Pan, J. Jing, and Y. Wang, Phys. Lett. B
\textbf{749}, 437 (2015) [ arXiv:1508.05926].


\end{thebibliography}
\end{document}